\documentclass{emulateapj}

\newcommand{\ls}{\raisebox{-0.5ex}{$\,\stackrel{<}{\scriptstyle                      
\sim}\,$}}

\shortauthors{Parker et al.}
\shorttitle{Galaxy-Galaxy Lensing in the CFHTLS}
\usepackage{graphics}
\tighten

\begin{document}

\def \farcs{\hbox{$.\!\!^{\prime\prime}$}}
\def \farcm{\hbox{$.\!\!^{\prime}$}}

\title{The Masses and Shapes of Dark Matter Halos from Galaxy-Galaxy Lensing in the
CFHTLS}

\author{Laura C. Parker} 
\affil{Department of Physics and Astronomy, University of Waterloo, Waterloo, ON, N2L 3G1, Canada}
\affil{European Southern Observatory, Karl-Schwarzschild-Str. 2, 85748 Garching, Germany}

\author{Henk Hoekstra}
\affil{Department of Physics and Astronomy, University of Victoria, Victoria, BC, V8W 2Y2, Canada}

\author{Michael J. Hudson}
\affil{Department of Physics and Astronomy, University of Waterloo, Waterloo, ON, N2L 3G1, Canada}

\author{Ludovic Van Waerbeke}
\affil{Department of Physics and Astronomy, University of British Columbia, Vancouver, BC V6T 1Z1, Canada}

\and

\author{Yannick Mellier}
\affil{Institut d'Astrophysique de Paris, 98bis, Boulevard Arago, 75014 Paris, France}

\begin{abstract}

We present the first galaxy-galaxy weak lensing results using early data
from the Canada-France-Hawaii Telescope Legacy Survey (CFHTLS). These
results are based on $\sim$22 deg$^2$ of $i^\prime$ data.  From this data, we estimate
the average velocity dispersion for an L* galaxy at a redshift of 0.3 to
be  137 $\pm$ 11 km s$^{-1}$, with a virial mass, M$_{200}$, of $1.1\pm 0.2$ $\times 10^{12}$ $h^{-1} $M$_\odot$ and a rest frame mass-to-light ratio of 173$\pm$34 $h$M$_\odot$/L$_{R_c\odot}$. We also investigate various possible sources of systematic error in detail.  Additionally, we separate our lens sample into two sub-samples, divided by apparent magnitude, thus average redshift. From this
early data we do not detect significant evolution in galaxy dark matter
halo mass-to-light ratios from a redshift of 0.45 to
0.27. Finally, we test for non-spherical galaxy dark matter halos. Our
results favor a dark matter halo with an ellipticity of $\sim$0.3 at the 2 $\sigma$ level when averaged over all galaxies. If the sample of foreground
lens galaxies is selected to favor ellipticals, the mean halo
ellipticity and significance of this result increase.

\end{abstract}

\keywords{gravitational lensing, dark matter, mass-to-light ratios, galaxy halos, halo shapes}

\section{Introduction}

It is widely accepted that galaxies live in massive dark matter halos, but the properties of these halos are not particularly well understood. Dark matter halos around individual galaxies introduce small coherent distortions to the shapes of background galaxies. This signal, weak gravitational lensing, can be used to infer properties of foreground galaxy dark matter halos such as their sizes and shapes \citep[e.g.,][]{bbs,hudson,henk03,sheldongg,henk04,henk05,mandelbaum06c,heymans06}.  Galaxy-galaxy lensing also provides an important link between numerical simulations, which model the dark matter very well, and other observational techniques, which are restricted to studying the (likely) biased luminous matter. The ability to connect observed galaxies to the properties of their dark matter halos provides important insights into the details of galaxy formation. Galaxy-galaxy lensing can also be used to constrain alternative theories of gravity. If observations such as flat rotation curves are due to a modified gravity law then the weak lensing signal at large distances from the galaxy center should be isotropic. If an anisotropic galaxy-galaxy lensing is observed around galaxies then this is a strong evidence for flattened dark matter halos and  disfavors any modified gravity theory \citep[e.g.][]{mortlock}.

Other techniques to map the dark matter content of galaxies, including extended rotation curves and dynamical methods, require visible tracer populations and are thus capable of probing only the inner regions of galaxy halos. Using satellite galaxies to probe the potential wells of galaxies offers an interesting alternative  \citep[e.g.,][]{zaritsky,prada,conroy}, but this technique requires the assumption of dynamical equilibrium and is susceptible to the inclusion of interlopers which can bias the results.

The study of galaxy-galaxy lensing has grown dramatically since its first detection 10 years ago \citep{bbs}, largely due to improved analysis techniques and a wealth of wide-field data. The basic requirement for a weak lensing measurement is wide-field imaging of reasonable depth and image quality. The most precise galaxy-galaxy lensing measurements to date are from the Red-Sequence Cluster Survey \citep{henk04} and the  Sloan Digital Sky Survey (SDSS) \citep{mandelbaum06c} which are not particularly deep surveys, but cover enormous areas. Measurements have been made from space in small fields \citep[e.g.,][]{hudson} and from large ground-based observations \citep[e.g.,][]{sheldongg,henk04,mandelbaum06c}. 

In this study we examine the weak lensing signal from a $\sim$22 square
degree area of sky from 2 fields of the WIDE component of the Canada France Hawaii
Telescope Legacy Survey (CFHTLS)\footnote{http://www.cfht.hawaii.edu/Science/CFHLS/}. These data represent a small fraction of the total CFHTLS-WIDE area which will be covered by the end of the survey in 2008 (approximately 170 square degrees). This early analysis is based on $i^\prime$ data only and is therefore lacking colors and redshift information for both the lenses and the sources.  Compared to previous large surveys used for galaxy-galaxy lensing \citep[e.g.,][]{henk04,sheldongg} the CFHTLS is deeper and therefore we have a higher source density and can probe galaxy halos at systematically higher redshifts.

In this paper we will introduce the CFHTLS data (\S{2}) and explain the redshift distribution determination and the galaxy shape measurements. In \S{3} we will show the first CFHTLS galaxy-galaxy lensing measurements including the determination of the average halo velocity dispersion as well as estimates for the average galaxy mass and mass-to-light (M/L) ratio. We also investigate a number of possible sources of systematic error. We will conclude this section by dividing the lens sample into a low and high redshift sample in order to look for evolution in the galaxy halos.  In \S{4} we will present a measurement of the halo shapes using anisotropic weak lensing. In \S{5} we will summarize and discuss these results.

Throughout the paper a cosmology with $\Omega_m=0.3$ and $\Omega_\Lambda=0.7$ is assumed. Results are presented in units of the hubble parameter, $h$ (the Hubble constant rescaled in units of 100 km s$^{-1}$Mpc$^{-1}$), which is assumed to be 1.

\section{Data}

Canada and France have united to use a large fraction of their telescope time at the Canada-France-Hawaii-Telescope (CFHT) in order to complete a 5-year photometric survey. The survey makes use of the new MegaCam instrument at CFHT. The camera is comprised of 36 separate CCD chips and produces distortion corrected 1 sq. degree images with superb image quality. The survey is divided into DEEP, WIDE and VERY WIDE components. The wide component is primarily designed for weak lensing studies. These data are well suited to the study of galaxy dark matter halos, as will be discussed in this paper, but also the study of lensing by the large scale mass distribution in the universe, cosmic shear \citep{cfhtlsshear}. Cosmic shear studies complement the Type Ia supernovae analysis from the CFHTLS-DEEP  survey \citep{cfhtlsdeep} since they can both be used to determine cosmological parameters, in particular the dark energy equation of state.

This galaxy-galaxy lensing project makes use of early data from the CFHTLS-WIDE survey. The wide data will eventually cover $\sim$170 sq. degrees in 5 filters ($u^{*}$,$g^{\prime}$,$r^{\prime}$,$i^{\prime}$,z$^{\prime}$). The observations are divided into 4 large patches which are well-separated in right ascension. Each patch is located far from the Galactic Plane in order to minimize extinction and contamination from bright stars. 

\begin{figure}
\plottwo{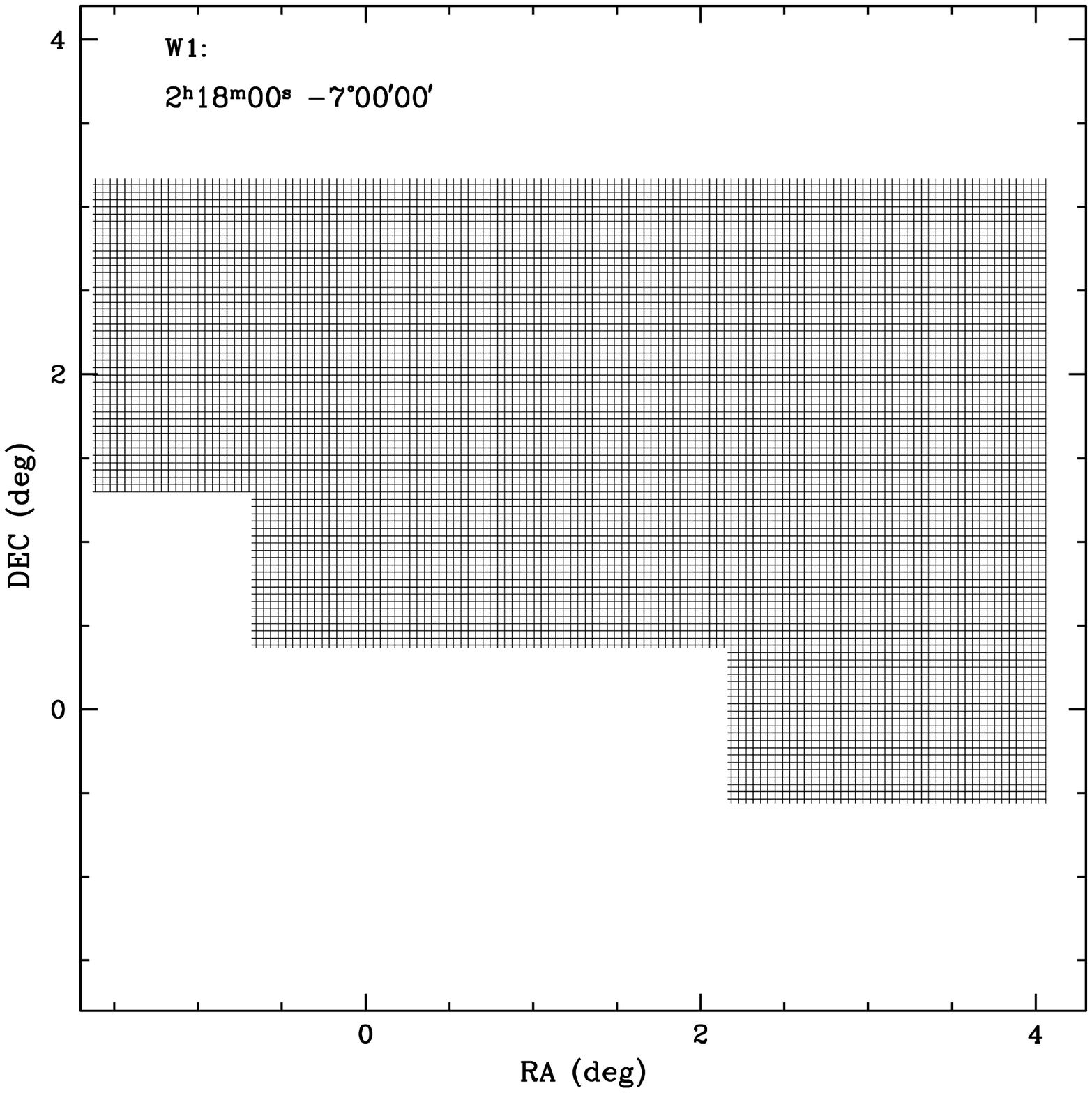}{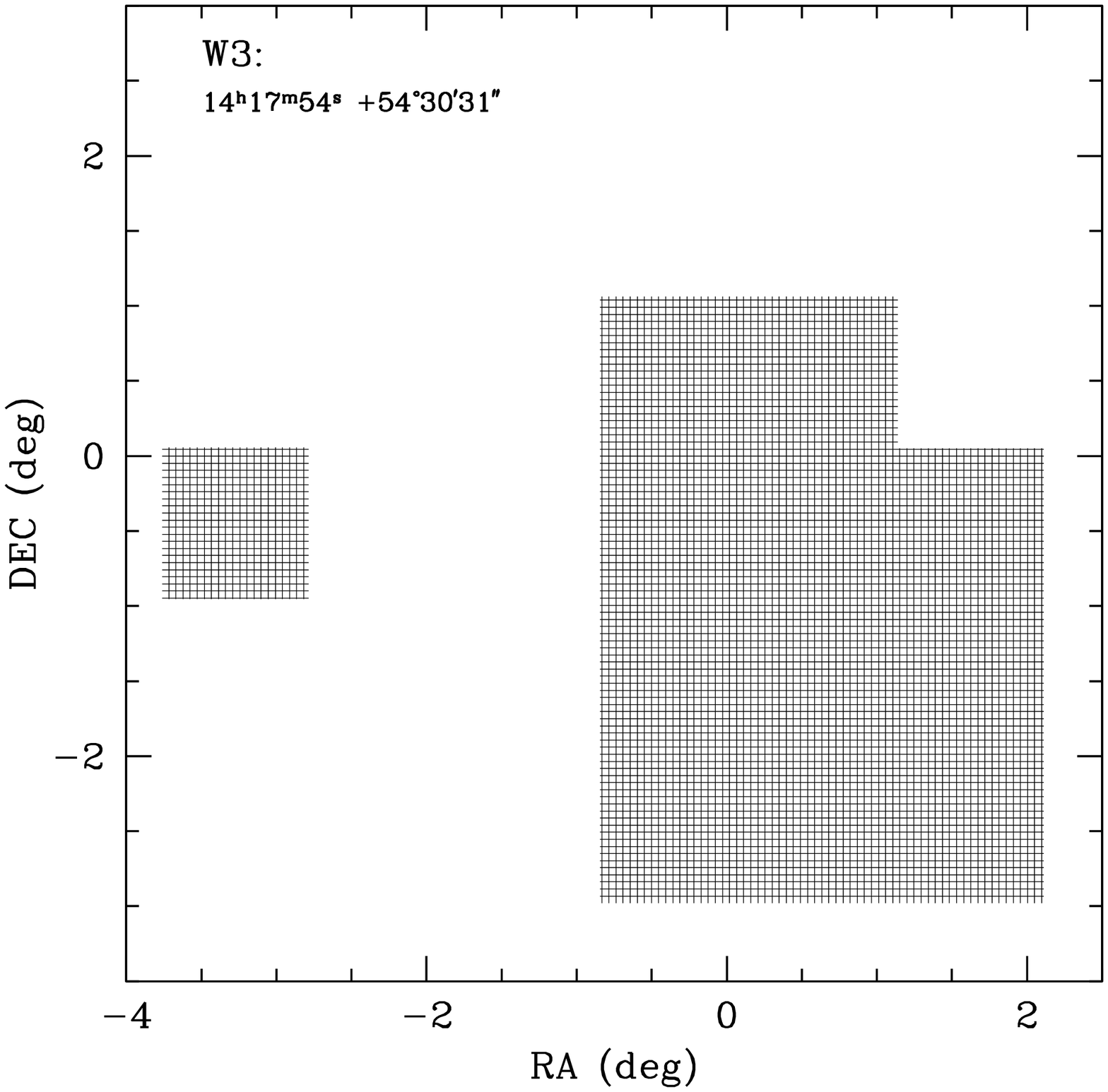}
\caption[CFHTLS W1 and W3 observations]{CFHTLS observations of W1 and W3 fields. The effective area is approximately 22 deg$^2$ from 31 pointings (19 deg$^2$ in W1 and 12 deg$^2$ in W3 ). Bright stars and bad pixels are masked.}\label{fig:data}
\end{figure}

The early data used in this paper study is based on 31 pointings taken in the $i^\prime$-band and covers an effective area of roughly 22 deg$^2$.   The data are from the W1 and W3 fields which will total 72 and 49 deg$^2$ respectively at the end of the survey. W1 is centered at 2$^{\rm{h}}$18$^{\rm{m}}$00$^{\rm{s}}$  -7$^{\circ}$00$^{\prime}$00$^{\prime\prime}$ and overlaps with the XMM-LSS field. W3 is centered at  14$^{\rm{h}}$17$^{\rm{m}}$54$^{\rm{s}}$  +54$^{\circ}$30$^{\prime}$31$^{\prime\prime}$ and overlaps with the Groth strip field. Figure \ref{fig:data} shows the coverage of the data used in this paper, totaling 19 deg$^2$ from W1 and 12 deg$^2$ from W3. The images are masked to avoid bright stars, diffraction spikes and bad pixels. The source density is $\sim$20 galaxies per sq. arcminute. These data were obtained in the 2003B, 2004A and 2004B observing semesters. The median seeing for the images used in this project is 0.76$^{\prime\prime}$, however,  more recent CFHTLS observations have improved image quality. The image quality degrades towards the edge of the images but only data with sub-arcsecond seeing are included in this analysis. More details on the data used in this analysis can be found in \cite*{cfhtlsshear}.

\subsection{Images and Catalogs}

The $i^\prime$ images used in this analysis are provided to the Canadian and French Astronomical communities via the Canadian Astronomical Data Centre (CADC). The basic data reduction pipeline, {\it Elixir}, is used to provide flat-fielded, de-biased images as well as photometric zeropoints and an initial astrometric solution.  The CFHTLS data are complicated by the fact that there are large dithers between some of the images. Therefore, some objects may appear on different chips in the individual exposures. This can lead to complicated PSF anisotropies, which are difficult to model and correct for. With this early analysis all such ``multi-chip''  data are excluded. Each chip is stacked separately with data only from that chip. This comes at a cost of losing approximately 20\% of the area but ensures accurate PSF anisotropy correction. Using stacks where all of the data comes from only one chip is a conservative approach and since the Megacam PSF is smooth from chip to chip future analyses will not impose this restriction. As described in \cite*{cfhtlsshear} any images with poor seeing ($>1$ arcsecond) are excluded from the stacks. Additionally, the images are inspected by eye to mask out areas contaminated by bleeding stars, diffraction spikes or other cosmetic defects. The stacks for this CFHTLS data were created using the SWarp routine\footnote{http://terapix.iap.fr/rubrique.php?id\_rubrique=49}. The stacks are created from 6 or 7 individual images of 620 second exposures, thus each stack is $>$1 hour and reaches a depth of $\sim$25 in $i^\prime$. 

A critical aspect of weak lensing measurements is accurately determining the shapes of faint objects. The ability to do this with stacked images requires very accurate astrometric solutions so that no spurious shape distortions are introduced in the stacking procedure. For this reason it is necessary to improve upon the basic astrometry from the {\it Elixir} processed data and include higher order corrections.  The images and catalogs used in this galaxy-galaxy lensing analysis are the same as those used in the first cosmic shear analysis of the CFHTLS-WIDE \citep{cfhtlsshear}.  The catalog details can be found in the shear paper, but we will outline the major points briefly here.

The basic technique for correcting the image astrometry is as follows:

 \begin{itemize}
 \item{Retrieve a red image from the second generation Digital Sky Survey (POS \rm{II}) for each pointing }
 \item{Use SExtractor \citep{bertin} to generate a high density catalogue of stars }
 \item{Match astrometric catalogue to each MegaCam image, averaging the positions of objects which appear on more than one chip (due to the large dither offsets)}
 \item{Use this master catalog to generate an astrometric solution for each chip}
 \end{itemize}
 
 Following the above procedure a stack can be created using all of the input images with the new astrometric solution applied. As was discussed above the stacks were created for each chip individually, and include only data from that chip, so that the PSF pattern could be corrected properly. This reduces the total area  but simplifies the PSF corrections. 
 
\subsection{Galaxy Shapes}

The carefully stacked $i^\prime$ images were processed with the peak finding algorithm of KSB \citep{ksb}. Objects which were detected to be more than 5$\sigma$ above the sky background were added to the source catalogue. The catalogue is then cleaned to include only objects larger than the PSF (points to the right of the stellar locus, identified in a plot of magnitude versus half-light radius as in \citet{henk02}). These objects are then analyzed in more detail in order to determine their apparent $i^\prime$ magnitudes and half-light radii, as well as their shape parameters. The shape is defined by two polarization ``vectors'' (equations \ref{eqn:e1} and \ref{eqn:e2}):

\begin{equation}
  e_1=\frac{Q_{11}-Q_{22}}{Q_{11}+Q_{22}}\label{eqn:e1}
\end{equation}

\begin{equation}
  e_2=\frac{2Q_{12}}{Q_{11}+Q_{22}}\label{eqn:e2}
\end{equation}

\noindent where $Q_{ij}$ are weighted with a Gaussian function that scales with galaxy size

\subsubsection{Shape Corrections}

The measured shapes must be corrected for distortions such as the effects of seeing and PSF anisotropy. This is done following the techniques outlined in \cite*{ksb} (KSB) and  \cite{henk98}. A sample of stars is found in the images and they are used to characterize the seeing and any anisotropy in the PSF. The shapes of the stars are fit with a second order polynomial for each chip in the CFHTLS stacks individually (that is to say, the stars are fit in the final stacked image, but each stacked chip is handled individually). The shapes of the stars are then used to correct the shapes of all of the galaxies. The catalogues include the position, magnitude, shape, error and P$^\gamma$ information, where P$^\gamma$ is the pre-seeing shear polarizability and can be calculated from the images of stars and galaxies \citep{henk98}. 

The procedure outlined here to correct the galaxy shapes was recently tested in the Shear TEsting Program \citep{heymans}, and with more realistic PSFs  in \cite{step2}. The results indicated that the procedure followed to create the catalogues in this analysis can reliably be used to measure weak shear to an accuracy within $\sim$2\%. It is important to note that the morphologies in the \cite{step2} and \cite{heymans} papers are simplified and do not consider possibilities such as shear and noise that vary as a function of position. The simulated images used in the \cite{step2} analysis are based on the best available imaging from {\emph{Suprime-Cam}} and thus do not have the identical signal-to-noise or PSF size as the CFHTLS images, therefore the quoted 2\% shear accuracy should be taken as a rough estimate. However the lack of B-modes in the cosmic shear analysis using these catalogues \citep{cfhtlsshear} is a powerful indicator that the level of systematics arising from PSF anisotropy is very small. Furthermore, galaxy-galaxy lensing is not very sensitive to PSF anisotropy as we average over many pairs of galaxies which have random directions of PSF anisotropy.

\subsection{Redshift Distribution}

\begin{figure}
\plotone{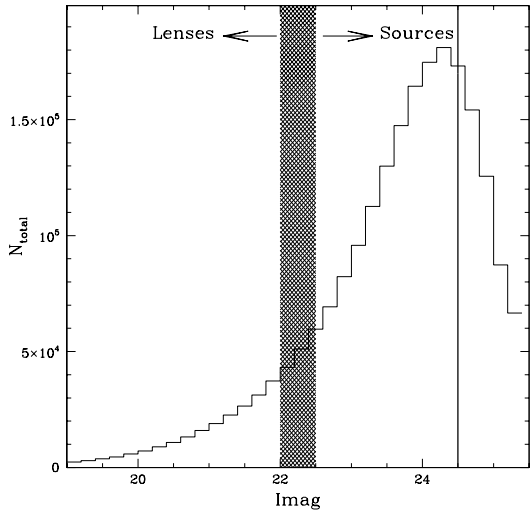}
\caption[CFHTLS-WIDE magnitude distribution]{Lens and source magnitude distributions. This histogram shows the distribution of $i^\prime$ magnitudes for all galaxies in the W1 and W3 fields used in this analysis. The hatched region divides the lenses and the sources, and the vertical line at $i^{\prime}$=24.5 shows the upper magnitude for the sources used.}\label{fig:maghist}
\end{figure}

The final CFHTLS-WIDE data set will include photometric redshifts for every faint source, but at present we have only single filter data available. 
For the analysis presented here, we select a sample of lenses and sources on the basis of their apparent $i^\prime$ magnitude. We define galaxies with $19<i^\prime<22$ as lenses, and galaxies with $22.5<i^\prime<24.5$ as sources which are used to measure the lensing signal. The faint end of the source distribution is chosen to ensure high signal-to-noise shape measurements and corresponds to the peak in the magnitude distribution (Figure \ref{fig:maghist}). This selection yields a sample of $\sim$2$\times 10^5$ lenses and $\sim$1.3$\times 10^6$ sources. These catalogues are used to generate 3.7$\times$10$^7$ lens-source pairs within a projected radius of 2 arcminutes of the stacked lenses. All lens-source pairs within 7 arcseconds of the host galaxy are eliminated from the catalogue since their shape measurements are likely compromised by light from the host lens. The angular scale of 7 arcseconds to 2 arcminutes corresponds to a physical scale of $\sim$25-500 h$^{-1}$kpc at the median redshift of the lenses.  Figure \ref{fig:maghist} shows the magnitude distribution for the entire sample of galaxies in the W1 and W3 fields used.

The lensing signal for an isothermal sphere is a function of $\langle\beta\rangle$, the average ratio of the angular diameter distances between the lens and source,D$_{LS}$, and between the observer and the source, D$_S$ as follows
\begin{equation}
  \beta=\rm{max}\left[0,\frac{D_{LS}}{D_S}\right], \label{eqn:beta}
\end{equation}
Therefore, in order to interpret the detected shear measurements it is necessary to know the redshifts of both the lenses and the sources. If the precise redshifts are not known for each object then at least their distributions must be understood in order to convert shear measurements into physical properties such as velocity dispersions and halo masses. The CFHTLS data used here were taken in a single band and so the redshift distributions of the sources and lenses must be estimated. The shear can only be estimated in projected angular bins and not physical units such as kiloparsecs. Therefore, the lensing signal for a distant galaxy is measured on a much larger physical scale than for the case of a closer galaxy. The mixing of scales complicates the interpretation of the results, but we can still learn about the average properties of halos.

The lens catalogues contain relatively bright galaxies, and therefore the distribution of the redshifts is quite well understood from previous studies, such as the CNOC2 field galaxy survey \citep{yee}. It is much more difficult to estimate the redshifts of faint sources, since they are generally too faint for spectroscopic redshift determination. The redshifts of faint galaxies are generally estimated using photometric redshifts. A spectroscopic study of the Hubble Deep Field (HDF) to depths of $\sim$24 in R$_{\rm{c}}$ showed that the spectroscopic redshifts agreed well with the photometric redshifts \citep{cohen}. 

\begin{figure}
\plotone{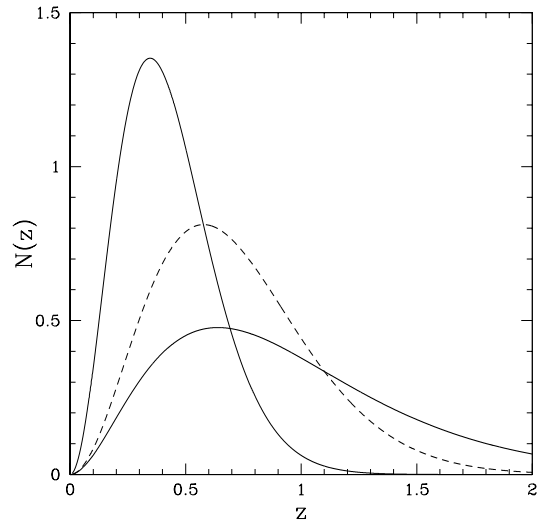}
\caption[N($z$) distribution]{N($z$) distribution. The lens and source distributions used in this analysis are shown with the solid lines. The sources have a median redshift of $\sim0.9$ and the lenses have a median  redshift of $\sim0.4$. The dashed line shows the redshift distribution of the sources based on the photometric redshifts from the HDFs \citep{fern}. The sources are at lower redshift using the HDF photometric redshifts, but the value of $\langle\beta\rangle$ is only $\sim$5\% smaller.}\label{fig:nz}
\end{figure}

The source and lens redshift distributions used in this analysis can be seen in Figure \ref{fig:nz}. The median lens redshift is $\sim$0.4 and the median source redshift is $\sim$0.9. The N($z$) distribution for the lenses was estimated using the functional form of  \cite*{brown} based on COMBO-17 data. The COMBO-17 survey uses a combination of 17 wide and narrow band filters to measure very accurate photometric redshifts in a 1 sq. degree patch.

\begin{equation}
  \frac{dN}{dz} \propto \frac{z^2}{z_*^3} {\rm{exp}} \left[-\left(\frac{z}{z_*}\right)^{1.5}\right]
\end{equation}

\noindent where $z_*$ is related to median redshift of the distribution by $z_*=z_m$/1.412, and $z_m$ is a function of the apparent magnitude. 

The source redshift distribution used in the analysis was estimated from photometric redshifts in the CFHTLS-DEEP fields \citep{ilbert}. The functional form of the N($z$)$_{\rm{sources}}$ was as follows

\begin{equation}
N(z)_{\rm{sources}}=r_{\rm{norm}} \left(\frac{z}{z_0}\right)^\lambda {\rm{exp}}\left[-\left(\frac{z}{z_o}\right)^\omega\right]
\end{equation}

\noindent where 

\begin{equation}
  r_{\rm{norm}}=\frac{\omega}{z_0 \Gamma\left(\frac{1+\lambda}{\omega}\right)}
\end{equation}

\noindent The values for $\lambda$, $\omega$ and $z_0$ were calculated using the magnitude cuts of our source catalogues applied to the \cite{ilbert} data.

The N($z$) distributions can then be used to estimate the angular diameter distances to the lenses and sources. Assuming a standard $\Lambda$CDM cosmology this results in a $\langle\beta\rangle$ of 0.49. If we alternatively used a source distribution based on the Hubble Deep Fields North and South \citep{fern} then we find a $\langle\beta\rangle$ of 0.46$\pm$0.02, where the error is from the field-to-field variations in the HDFs, and the finite number of galaxies. The $\beta$ estimates are weighted in the same manner as the shear estimates as will be discussed in the next section. The difference is source redshift distributions between the CFHTLS-DEEP fields and the HDFs is obvious in Figure \ref{fig:nz}. The sources are at slightly higher redshifts when using the CFHTLS photometric redshifts than with distribution estimated from the HDF. The visible discrepancy between the photometric redshift distributions from the CFHTLS and the distributions based on the HDF is not statistically significant when one accounts for uncertainties in photometric redshift distributions due to cosmic variance affecting small fields (see Fig 1 of  \cite{vw06} for a dramatic illustration of this in the HDF). The different redshift distributions only change the $\langle\beta\rangle$ value by $\sim$5\%.

\section{Analysis}

The tangential shear signal must be fit with an assumed halo mass model in order to extract physical properties of the halo such as velocity dispersion and mass. A common mass model assumed is the isothermal sphere for which the tangential shear is proportional to the Einstein radius and hence to the velocity dispersion squared (equations \ref{eqn:ein2} and \ref{eqn:gt}).

 \begin{equation} 
    \theta_E=\left(\frac{4\pi\sigma^2}{c^2}\right)\beta\\    	
    =\left(\frac{\sigma}{186\rm{km s}^{-1}}\right)^2\beta \hspace{1mm} [^{\prime\prime}]\label{eqn:ein2}
  \end{equation}

\begin{equation}
\gamma_T=\frac{\theta_E}{2\theta}=\frac{2\pi\sigma^2}{c^2 \theta }\beta\label{eqn:gt}
\end{equation}
  
\begin{figure}
\plotone{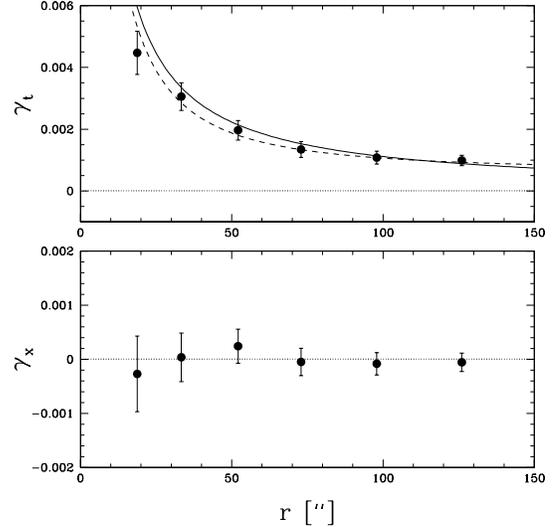}
\caption[Tangential and cross-shear in angular bins around CFHTLS galaxies]{(a) The ensemble averaged tangential shear as a function of radius around a sample of CFHTLS galaxies with 19$\ls$$i^\prime\ls$22. The best fit isothermal sphere, shown with the solid line, yields an Einstein radius of
  0{\farcs}210$\pm$0{\farcs}017, corresponding to a velocity dispersion of 132$\pm$10 km s$^{-1}$. The dashed line represents the best fit NFW profile, corresponding to a halo with an r$_{200}$ of 150 h$^{-1}$kpc.  (b) The signal when the source images are rotated by 45$^\circ$. Note the difference in scales between the top and bottom figures. No signal is present as expected if the signal in (a) is
  due to gravitational lensing.}\label{fig:cfhtlset}
\end{figure}

In this study the lenses are stacked together and the sources that lie within a projected radius of 2 arcminutes are divided into angular bins. The component of their shape tangential to the lens center is determined and averaged in each bin. Each shear calculation is weighted by the error in the shape measurement as described in Hoekstra et al. (2000). Galaxy-galaxy lensing measurements also have a convenient built-in systematic test. If the tangential lensing signal detected is due to gravity then it should vanish if the source images are rotated by 45 degrees. The tangential and ``cross-shear'' for the entire sample are plotted in Figure \ref{fig:cfhtlset}. The best fit isothermal sphere has an Einstein radius of 0{\farcs}247$\pm$0{\farcs}020. A lensing signal is detected at high significance ($>12\sigma$). The cross-shear measurement is consistent with 0, as expected, and therefore the tangential shear signal is interpreted as being caused by weak lensing from an isothermal sphere potential. 

The tangential shear measurements can also be fit with other dark matter profiles, such as the Navarro, Frenk and White (NFW) profile \citep{nfw}, given by

\begin{equation}
\rho(r)=\frac{\delta_c\rho_c}{(r/r_s)(1+r/r_s)^2}\label{eqn:nfw}
\end{equation}

\noindent where $\rho_c$ is the critical density for closure of the universe. The scale radius, $r_s$, is defined as $r_{200}/c_{NFW}$ where $c_{NFW}$ is the dimensionless concentration parameter, and $\delta_c$ is the characteristic over-density of the halo. The tangential shear equations for a NFW halo can be found in \cite{bart} and \cite{wrightbrain}. The NFW density profile has 2 free parameters, but in this analysis we will consider the concentration parameter to be fixed, obeying the relation between virial mass and concentration found in simulations by Bullock et al. (\citeyear{bullock}). The best fit NFW profile, corresponding to a halo with r$_{200}$ of 0.15h$^{-1}$Mpc (M$_{200}$ of 7.6$\times$10$^{11}$h$^{-1}$M$_{\odot}$) is shown as the dashed line in Figure \ref{fig:cfhtlset}.

\subsection{Systematics}

In order to verify the validity of our shear signal we have performed a number of systematic tests. At the outset we are confident that the catalogue used in this analysis is free of major systematics because there were no significant B-modes found in the cosmic shear analysis using the same catalogue \citep{cfhtlsshear}. The initial systematic test is to calculate the cross-shear, as discussed above. We have also considered the fact that some lenses and sources are physically associated. We can estimate the level of this association by calculating the number density of galaxies in each bin, which decreases with radius.  We have estimated the fractional excess of background sources around lenses, $f_{bg}$, and increased the shear signal by $1+f_{bg}$ to compensate. The background density of sources was estimated by counting the number of sources in bins around random lens positions. The random lens catalogue was the same size as the lens sample used in this analysis.  Our best fit $f_{bg}=0.82r^{-0.63}$, where $r$ is the radius in arcseconds, is similar to what has been found in previous measurements \citep[e.g.][]{henk04}  The fractional excess is a strong function of radius and thus only the inner angular bins are influenced by physically associated lens-source pairs. This correction has been applied to all quoted values in this paper (e.g.,  $\theta_E$, $\sigma$), and was calculated for each lens sample separately.

\begin{figure}
\plotone{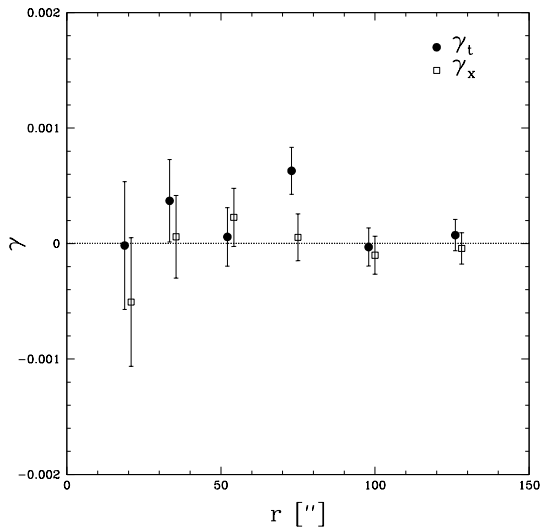}
\caption[Tangential and cross-shear in angular bins around random points]{The ensemble averaged tangential (filled circles) and cross (open squares) shear as a function of radius around a sample of random points in the CFHTLS fields. Both the signals are consistent with zero as expected. Note that the cross-shear points have been slightly shifted so that the error bars are clearly visible}\label{fig:randshear}
\end{figure}

A further test of the signal can be made by measuring the tangential and cross-shear around random points in the field. The results from performing this test are demonstrated in Figure \ref{fig:randshear} and are consistent with no detected signal, as expected. The same number of random points were used as the number of foreground lens galaxies.

It is also important to examine the influence of intrinsic alignments. Without redshift information for our lenses and sources it is possible that physically associated lens-source pairs are included in the analysis. If these pairs are aligned then the tangential shear signal will be contaminated. Recent evidence has suggested that satellite galaxies' major axes preferentially point towards their host galaxies \citep{agustsson1} which would decrease the galaxy-galaxy lensing signal. If we adopt the most pessimistic view and consider a very large contamination in the inner 250 kpc  due to intrinsic alignment, we can compensate by boosting the tangential shear signal. If we apply a 40\% boost to the inner 3 bins in Figure \ref{fig:cfhtlset}  (corresponding to roughly 250 kpc at the redshift of our lenses), adopting the worst case from Agustsson \& Brainerd (25-40\% suppression of shear on scales less than 250 kpc), then the best fit isothermal sphere yields an Einstein radius which is 7\% larger than we have quoted above. This small difference is  due to the fact that the outer 3 data points, with smaller errorbars, constrain the isothermal sphere fit much more than the 3 inner bins. This is an upper limit on the contamination from intrinsically-aligned satellites. 

Alternatively, we can also estimate the influence of intrinsic alignment by repeating the analysis after dividing the distant sources into 2 sub-samples based on their apparent magnitudes, while keeping the lens sample identical. After re-scaling by the new $\beta$ values the shear signal should be the same since the lens sample has not changed. If, however, there is significant intrinsic alignment, then the signal from the brighter sources (lower redshift) will be suppressed. We have carried out this test and found that the velocity dispersion derived from the shear signal computed using the brighter sources is indeed suppressed by $\lesssim$10\%. This suppression due to intrinsic alignments for the entire sample would therefore be $<$10\%. We have not applied any correction for intrinsically-aligned sources, instead we include this as part of our total systematic error budget.

One final potential systematic that should be discussed is possible contamination from stellar objects in the final source catalogue. Moderately bright stars are easily identified as they lie in a tight vertical locus in plot of magnitude versus size. Bright stars saturate and their measured sizes increase, but they can also be easily removed from the catalogues. There is some overlap between faint stars and galaxies  which can dilute the shear signal, but by extrapolating the number counts of stars to faint magnitudes it is clear that stars will contribute at most a few percent to the total number counts. In addition, we consider only sources larger than the stellar PSF which cleanly separates stars and galaxies to $i^\prime \sim 24$ \citep{henk02}. Furthermore, in this analysis we down-weight faint sources which significantly reduces the contamination from stars (Heymans et al. 2006). The weight for sources fainter than $i^\prime$ of 24 is very low, thus the level of stellar contamination is estimated to be $<$1\%. It is important to note that for the discussion of halo shapes below that this potential source of systematic error is irrelevant because stars are randomly distributed and the measurement is a relative one.

Based on the above considerations we estimate that the total systematic error in our measurements is at most 10\% (1-$\sigma$ level). The exact level of possible systematic errors is difficult to estimate as the error distributions for the various systematics are not known. Throughout the paper the quoted errors are always statistical errors derived based on the errors in the shape measurements and the number of sources in each bin. 
   
\subsection{Velocity Dispersion}

The best fit isothermal sphere yields a velocity dispersion of 132$\pm$10 km s$^{-1}$ (equations \ref{eqn:ein2} and \ref{eqn:gt}).  This measurement takes into account the fact that some lenses and sources will be physically associated. We can estimate the level of this association by calculating the number density of galaxies in each bin which decreases with radius.  

The velocity dispersion depends on the sample of lenses used and must be scaled in order to compare to other results. This can easily be done by assuming a scaling relation between luminosity and velocity dispersion as shown in equation \ref{eqn:lstar}.

\begin{equation}
  \frac{\sigma}{\sigma_*}=\left(\frac{L}{L_*}\right)^{1/\alpha }\label{eqn:lstar}
\end{equation}

\noindent where $\sigma_*$ is  the velocity dispersion of an L$_*$ galaxy. The scaling factor $\alpha$ is generally assumed to be 3 or 4, motivated by the observations of the Faber-Jackson relation \citep{fb}, for example. We based our L$_*$ galaxy on the CNOC2 luminosity function results \citep{lin}. Lin et. al measure L$_*$ at z$=0.3$, which after correction to the $\Lambda$CDM cosmology assumed here gives $1.3 \times10^{10}h^{-2}$L$_{\odot R_{\rm{c}}}$. We used the color and k-corrections from \cite*{frei} to convert our magnitudes to $R_{\rm{c}}$ and evaluate the luminosity of our lenses. All luminosities are calculated assuming the proportion of early and late types of galaxies found in the CNOC2 redshift survey, since the redshifts and brightnesses of the two samples are comparable (28\% early types, 24\% intermediate types, 47\% late types). It is also important to note that the average luminosity is calculated using the same weights used in the shear analysis.

The Einstein radius for an L$_*$ galaxy is given by 

\begin{equation}
  \theta_E=\frac{4\pi}{c^2}\frac{\sigma_*^2}{L_*^{2/\alpha}}\langle\beta L^{2/\alpha} \rangle
\end{equation}

\noindent where $\sigma_*$ is the velocity dispersion for an L$_*$ galaxy and $\theta_E$ is in radians. We estimate the average luminosity for our lens galaxies to be $\langle L \rangle=$1.1$\times$10$^{10}h^{-2}$L$_{\odot R_{\rm{c}}}$. The scaled velocity dispersion can now be  estimated for different assumed $\alpha$ values. The results of scaling the observed velocity dispersion for our sample of lenses to a typical L$_*$ galaxy are summarized in Table \ref{tab:sig}.

Note that, in principle, the tangential shear signal is due not only to the lens galaxy in question (the 1-halo term, assuming the galaxies are central galaxies), but also other mass clustered with the primary galaxy (the 2-halo term). In practice, though the velocity dispersion is estimated from the best-fit isothermal sphere out to a radius of $\sim$130$^{\prime}$$^{\prime}$, which at these redshifts corresponds to a physical scale of much less than 1 h$^{-1}$Mpc which should still be dominated by the 1-halo term \citep{zehavi}. This is not strictly true for satellite galaxies where the contribution to the lensing signal from the host halo would kick-in on smaller scales \citep{yoo}. However, as the shear profile is well fit by both Singular Isothermal Sphere and NFW profiles this is not a significant concern.

A galaxy-galaxy lensing analysis of the COMBO-17 data by Kleinheinrich et al. (2005) estimated the value for $\sigma_*$ for various scenarios, including when the redshift is known for every lens and every source, and when there is no redshift information available. The latter case can be compared with our results. They obtained a mean $\sigma_*$ of 156$^{+24}_{-30}$ km s$^{-1}$. Hoekstra et al. (2005) used data from the Red-Sequence Cluster Survey (RCS) to estimate the properties of galaxy dark matter halos using galaxy-galaxy lensing without redshift information. They found a value for $\sigma_*$ of  136$\pm{8}$ km s$^{-1}$. It is important to note that the estimate for L$_*$ is slightly different in each of these studies, but the results agree within the errors.

\subsection{Halo Masses}

The total extent and mass of dark matter halos can be estimated assuming a mass model for the galaxy halos as suggested by  \cite*{bbs} with a density profile

\begin{equation}
  \rho(r)=\frac{\sigma^2s^2}{2\pi Gr^2(r^2+s^2)}
\end{equation}

\noindent where $s$ is a measure of the truncation scale of the halo. This profile is an isothermal sphere at small radii with a cut-off at large radii, characterized by the scale $s$, which scales with velocity dispersion \citep{rix}

\begin{equation}
  s=s_*\left(\frac{\sigma}{\sigma_*}\right)^2
\end{equation}

\noindent Assuming this truncated isothermal sphere halo model, the mass enclosed within a sphere of radius $r$ is 
\begin{equation}
  M(r)=\frac{2\sigma^2s}{G} {\rm arctan} (r/s)
\end{equation}

\noindent which, because of the truncation, results in a finite mass \citep{henk04}
\begin{equation}
  M_{\rm{tot}}=\frac{\pi\sigma^2s}{G}=7.3\times10^{12}h^{-1}M_{\odot}\left(\frac{\sigma}{100 \rm{km s} ^{-1}}\right)^2\left(\frac{s}{1 \rm{Mpc}}\right)
\end{equation}

The truncation scale can be assumed to be constant for all halos, in which case the M/L ratio would be $\propto$ L$^{1/2}$, if L $\propto\sigma^4$. This is the favoured model in recent analyses \citep[e.g.][]{guzik, henk05}. Alternatively, it can be assumed that M/L is constant for all galaxy halos in which case $s \propto\sigma^2$ \citep{bbs,hudson}. We do not explore the various possible M/L scalings in this paper as we assume a value for the truncation radius from the literature and do not directly fit a truncated isothermal sphere model.

Assuming the truncation radius found by  \cite*{henk04} for an L$_{*}$ galaxy of 185$\pm$30 kpc we estimate the total mass of our L$_*$ galaxy to be $2.2\pm 0.4$ $\times 10^{12}$ $h^{-1} $M$_\odot$ if L$\propto\sigma^4$, and 2.4$\pm 0.5$ $\times 10^{12}$ $h^{-1} $M$_\odot$ if L$\propto\sigma^3$. The results are in good agreement with the results from the Red-Sequence Cluster Survey \citep{henk04}, which found total halo mass to be 2.7$\pm 0.6$ $\times 10^{12}$ $h^{-1}$M$_{\odot}$. We estimate the virial mass, M$_{200}$, of our galaxy halos, assuming an isothermal sphere model, to be 1.1$\pm 0.2$ $\times 10^{12}$ $h^{-1} $M$_\odot$, where M$_{200}$ is the mass inside the virial radius r$_{200}$ defined as the radius where the mass density is 200 times the critical density. This is in good agreement with the results from RCS where the virial mass of L$_*$ halos was estimated at $11.7\pm1.7\times 10^{11}$ $h^{-1} $M$_\odot$ \citep{henk05}. Our results are also in good agreement with results from the lower redshift SDSS sample where the virial mass of L$_*$ halo was found to be $13.2^{+6.0}_{-5.6}\times 10^{11}$h$^{-1}$M$_\odot$ \citep{mandelbaum06c}. In order to make this comparison we chose the central virial mass found for $<$L$/$L$_*$$>=$1.1 (see Table 3 of \citet{mandelbaum06c}) and scaled this to L$_*$. We then scaled the SDSS from an NFW with a virial radius defined using $\rho$=180$\bar{\rho}$ to a singular isothermal sphere profile with a virial radius defined by $\rho$=200$\rho_{crit}$. . The GEMS team has estimated the virial mass of halos for a sample at higher redshift \citep{heymans06}. They estimate that the virial mass for an L$_*$ galaxy at a redshift of $\sim0.65$ is $14.1^{+3.7}_{-4.5}\times 10^{11}$h$^{-1}$M$_\odot$. For this comparison we once again transformed from an NFW to a singular isothermal sphere profile, for an L$_*$ galaxy, using our definition of the virial radius. Our measured velocity dispersion for an L$_*$ galaxy is also in agreement with results from fundamental plane measurements \citep{sheth}. 

We can use the luminosity of an L$_*$ galaxy together with the mass estimates above  in order to calculate a typical M/L ratio. This leads to a M/L ratio for an L$_*$ galaxy of 173$\pm$34 $h$M$_\odot$/L$_{R_c\odot}$, assuming $\alpha=4$. 

\subsection{Evidence for evolution?}

The CFHTLS data are ideally suited to looking for evolution in dark matter halos with redshift. The high signal-to-noise of our primary result permits the division of the lenses into sub-samples at different redshifts. With this early data there is sufficient signal-to-noise to divide the foreground lens samples into multiple bins, however without redshift or color information this proves difficult. Nevertheless, as was indicated earlier we can  divide the lenses based on their observed $i^\prime$ magnitudes. The brighter lenses will be on average at a lower redshift while faint lenses will be, on average, at higher redshift.  We decided to apply a magnitude cut at  20.5 in $i^\prime$. This leads to a ``bright lens sample'' containing $\sim 4\times10^4$ galaxies and a ``faint lens sample'' containing $\sim 1.5\times10^5$ galaxies. This selection results in a low redshift sample with a median redshift of 0.27 and a high redshift sample with median redshift of 0.45.

\begin{figure}
\plotone{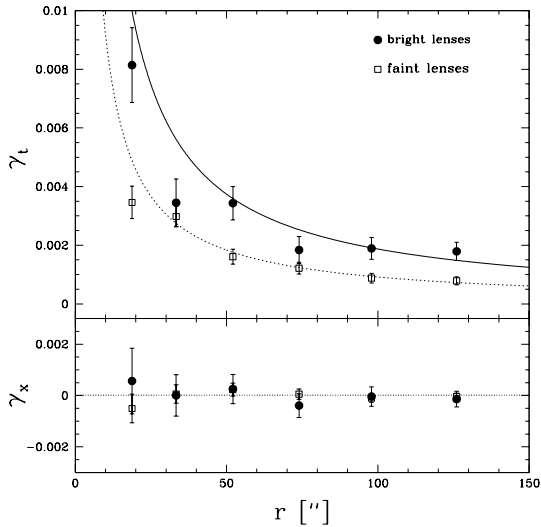}
\caption[Tangential in 2 redshift bins]{(a)The tangential shear signal for two samples of foreground lenses. (b) The cross-shear signal when the source images are rotated by 45$^\circ$. The faint lenses (indicated by open squares) have $i^{\prime}>20.5$ while the bright lenses (indicated by filled circles) have $i^{\prime}<20.5$. }\label{fig:bothshear}
\end{figure}

We repeated the tangential shear analysis described in the previous section and the results are shown in Figure \ref{fig:bothshear}. There is a striking difference in the tangential shear profiles for the two lens sub-samples.  However, as indicated in equation \ref{eqn:ein2}, the velocity dispersion is also dependent on the parameter $\beta$. The source catalogues used are the same as in the analysis of the entire sample, and therefore so is the source redshift distribution. However, these two samples of lenses have very different redshift distributions, thus their average $\beta$ values will also be different.  The faint lens population has a $\langle \beta\rangle$$=$ 0.44 while $\langle \beta\rangle$$=$0.67 for the bright lenses.

 \begin{table*}
\begin{center}
\begin{tabular}{|c|c|c|c|c|c|c|c|c|c|}
\hline
Sample &$\langle$z$\rangle$& $\langle\beta\rangle$ & $\langle$L$\rangle$ & $\langle\sigma^2\rangle^{1/2}$ & $\alpha$ & $\sigma_*$ & M$_{\rm{total}}$ & M$_{200}$& M/L$_*$\\

& & &10$^{10}h^{-2}$L$_{R_c\odot}$ & km s$^{-1}$ & & km s$^{-1}$   & 10$^{12}$$h^{-1} $M$_\odot$  & 10$^{12}$$h^{-1} $M$_\odot$ & $h$M$_\odot$/L$_{R_c\odot}$\\
\hline

Full & 0.35 & 0.49 &1.11 & 132$\pm$10& 4 & 137$\pm$11 & 2.2$\pm$0.4 &  1.1$\pm$0.2 &  173$\pm$34\\
Full & 0.35 & 0.49 & 1.11 & 132$\pm$10 & 3 & 141$\pm$12 & 2.4$\pm$0.5 &  1.2$\pm$0.2  & 189$\pm$37\\

Faint & 0.45 & 0.44 & 1.07 & 129$\pm$11 & 4 & 134$\pm$12  & 2.1$\pm$0.5 & 01.0$\pm$0.2 & 165$\pm$43\\

Faint & 0.45 & 0.44 & 1.07& 129$\pm$11 & 3 & 137$\pm$12  & 2.1$\pm$0.5 & 1.1$\pm$0.2  & 174$\pm$46\\

Bright & 0.27 & 0.67 & 1.26&  141$\pm$18  & 4 & 142$\pm$18 & 2.7$\pm$0.6 & 1.2$\pm$0.3  & 206$\pm$54\\

Bright & 0.27 & 0.67 & 1.26& 141$\pm$18  & 3 & 142$\pm$18  & 2.7$\pm$0.4  & 1.2$\pm$0.3  & 206$\pm$54\\

\hline
\end{tabular}
\caption{Properties for the various lens samples.  The quoted errors do not include mass model uncertainties.\label{tab:sig}}
\end{center}
\end{table*}

The differences in $\beta$ can explain most of the offset between the 2 shear profiles in Figure \ref{fig:bothshear}, however it is still interesting to scale the velocity dispersions to a typical L$_*$ galaxy to see if any real difference can be seen in the dark matter halos of the two lens samples. The results are shown in the Table \ref{tab:sig}. We do not see conclusive evidence for evolution in galaxy dark matter halos in this redshift range, but this is not so surprising due to the lack of photometric or spectroscopic redshift information to clearly divide the lens populations. 

\section{Halo Shapes}

An important insight into the nature of dark matter comes from the shapes of dark matter halos. Dynamical measurements can be used to trace out the halo shapes on small scales, but they can not be used on larger scales where there are no visible tracers. Numerical simulations of CDM indicate that dark matter halos should be flattened, and more often prolate than oblate \citep{dubinski,springel04,allgood}. Simulations of self-interacting dark matter produce more spherical halos \citep{dave}. 

Alternative theories of gravity such as Modified Newtonian Dynamics (MOND) \cite{milgrom2} attempt to explain astrophysical observations by modifying gravity rather than invoking non-baryonic dark matter. Non-relativistic theories of modified gravity can not provide predictions for lensing measurements, and therefore can not be tested by lensing observations. However, there is now a candidate relativistic modified gravity theory as presented by  \cite*{bek} which can be used as an alternative to dark matter to explain relativistic phenomena such as lensing. Thus far this theory appears to match many observations \citep{skordis,zhao}. One interesting test for this theory is the inferred shapes of galaxy dark matter halos from weak lensing \citep[e.g.][]{mortlock}.

Modified gravity theories predict that the lensing signal is due to the observed luminous material, and thus any anisotropy in the lensing measurement is due to the anisotropy in the distribution of gas and stars. Therefore, on small scales one would expect the lensing signal to be anisotropic since galaxies themselves are, but on the large scales probed by galaxy-galaxy lensing one would expect a nearly isotropic signal since there is no luminous material present at large radii from the galaxy center (provided that the galaxy is isolated and not in a group or cluster). If a highly anisotropic signal is detected at large radii this provides an interesting constraint on modified gravity theories, and provides supporting evidence for dark matter theories. The predictions from MOND of spherical halos assumes that the lens galaxies are isolated, and not in groups or clusters where there could be light contamination on large scales. 
 
Galaxy-galaxy lensing measurements have generally assumed that dark matter halos are spherical, but a recent galaxy-galaxy measurement by  \cite*{henk04} detected a significant flattening of dark matter halos. This result was not observed in the latest analysis of SDSS data by \cite{mandelbaumshape}.  However, Mandelbaum et al. did find marginal evidence for flattened halos when they restricted their lens sample to red galaxies. 

One important note is that all such measurements rely on the assumption that the mass distribution of the halo is aligned with the light distribution of the galaxy. If the halo flattening is not correlated with the light profile orientation then this simple measurement is much more difficult to interpret, and the flattening signal is likely systematically suppressed. 

Brainerd \& Wright (\citeyear{brainerd}) discussed how to measure halo shapes by looking for an anisotropic galaxy-galaxy lensing signal. One simple approach to try and detect if galaxy dark matter halos are non-spherical is to test whether the tangential shear signal is different along the semi-major and semi-minor axes of the visible galaxy. 

\begin{figure}
\plotone{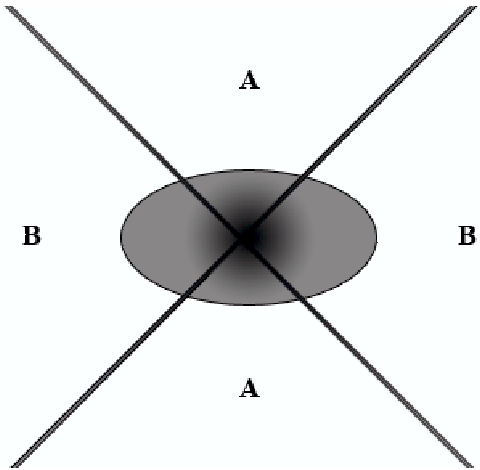}
\caption[Schematic diagram of anisotropic lensing]{Schematic of anisotropic shear. If galaxy halos are not spherical then there should be a difference in the tangential shear signal coming from the regions labeled with an A versus those labeled with a B.}\label{fig:ellipse}
\end{figure}

\begin{figure}
\plotone{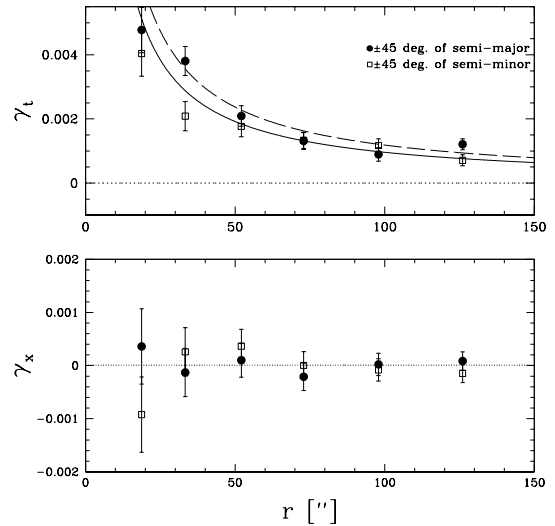}
\caption[Anisotropic tangential shear measurement]{Mean tangential shear for sources close to the semi-major axes (filled circles) and close to the semi-minor axis (open squares). The best fit isothermal sphere for the sources within 45$^\circ$ of the semi-major axes, as indicated by the dashed line, yields an Einstein radius of 0{\farcs}228$\pm$0{\farcs}026 corresponding a velocity dispersion of 138$\pm$15 km s $^{-1}$. The best fit isothermal sphere for the sources within 45$^\circ$ of the semi-minor axes, as indicated by the solid line, yields an Einstein radius of 0{\farcs}184$\pm$0{\farcs}018 corresponding a velocity dispersion of 122$\pm$12 km s $^{-1}$. The cross-shear is consistent with 0 as expected.}\label{fig:angshear}
\end{figure}

The analysis described in the previous sections was repeated for the galaxy lenses, this time dividing the sources into those within 45 degrees of the semi-major axis and those within 45 degrees of the semi-minor axis (see the schematic in Figure \ref{fig:ellipse}). The tangential shear results can be seen in Figure \ref{fig:angshear}. The signal from the two angular bins of sources is very similar. The best fit isothermal spheres yield Einstein radii of 0{\farcs}211$\pm$0{\farcs}021 and 0{\farcs}270$\pm$0{\farcs}031, corresponding to velocity dispersions of 122$\pm$12 km s $^{-1}$ and  138$\pm$15 km s $^{-1}$, respectively.
 
\begin{figure}
\plotone{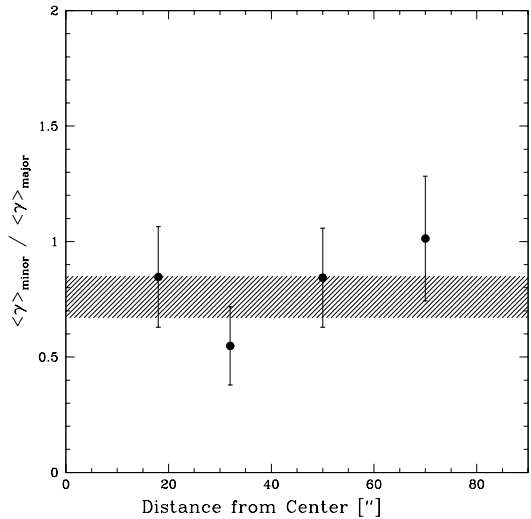}
\caption[Tangential shear ratio]{Ratio of mean shear experienced by sources closest to minor axes of a foreground lens to those experienced by sources closest to the major axes. The weighted average shear ratio is 0.76 $\pm $0.10 favoring a halo with an ellipticity of 0.3.}\label{fig:ourshape}
\end{figure}

We calculated the ratio of $\langle\gamma\rangle_{\rm{minor}}$ to  $\langle\gamma\rangle_{\rm{major}}$ out to 70$^{\prime}$$^{\prime}$ (which corresponds to $\sim$250 h$^{-1}$ kpc, for comparison to Brainerd \& Wright) and the results are plotted in Figure \ref{fig:ourshape}. The best fit shear ratio is 0.76$\pm$0.10,   indicating a $\sim$2$\sigma$ detection of non-sphericity for dark matter halos. Examination of the minor-to-major axis shear ratio as a function of radius in Figure 1 of Brainerd \& Wright (2000) suggests that a ratio of 0.76 corresponds to a halo ellipticity of $\sim$0.3. 

\begin{figure}
\plotone{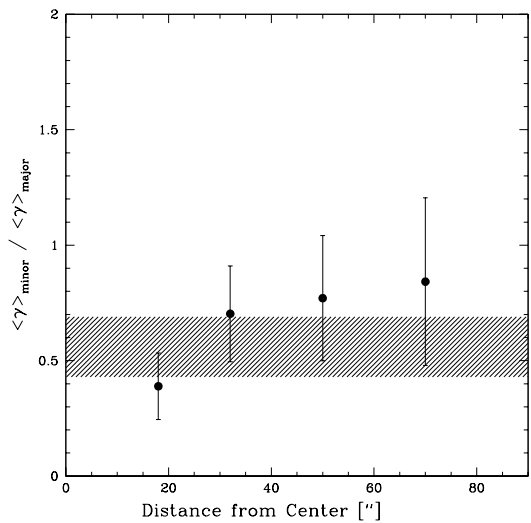}
\caption[Tangential shear ratio for elliptical lenses]{Ratio of mean shear experienced by sources closest to minor axis of lenses to those experienced by sources closest to the major axes as in Figure \ref{fig:ourshape} only with the roundest lenses (e$<0.15$ excluded). Lenses which appear circular on the sky have more poorly determine semi-major axes positions and therefore have been excluded from this plot. The weighted average shear ratio is 0.56$\pm$0.13.}\label{fig:ourshape2}
\end{figure}

It is important to measure the anisotropic weak lensing signal for a sample of well-understood lenses. A way to improve the measurement of anisotropic weak lensing is to select a sample of lenses with a well defined semi-major axis direction. Therefore it is perhaps prudent to choose only the brighter lenses, or discard those galaxies with noisy shape measurements, or with very little ellipticity. If a lens has an extremely small measured ellipticity, then determining an accurate semi-major axis location is very difficult and prone to errors.  We choose to repeat the anisotropic weak lensing signal for all lenses with an ellipticity $>$0.15 which corresponds to an axis ratio $<$0.8. The results are shown in Figure \ref{fig:ourshape2}. The best fit shear ratio is 0.56$\pm$0.13 which according to Brainerd \& Wright favors  halo model with an ellipticity of $\sim$0.5, although directly comparing to Brainerd \& Wright is not strictly correct as their results are based on simulations including the full distribution of galaxy shapes.

\begin{figure}
\plotone{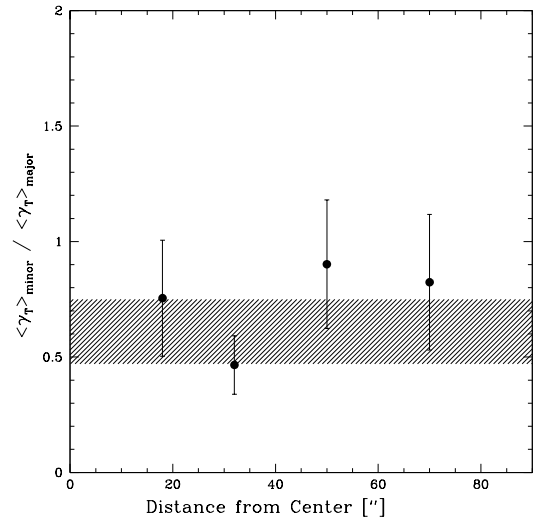}
\caption[Tangential shear ratio for early type galaxies]{Ratio of mean shear experienced by sources closest to minor axis of a lens to those experienced by sources closest to the major axes as in Figure \ref{fig:ourshape} only including just the lenses with axis ratio (b/a) between 0.5 and 0.8. This selection criteria will result in a lens sample with is dominated by early type galaxies. The weighted average shear ratio is 0.61$\pm$0.14.}\label{fig:ourshape3}
\end{figure}

In addition, it is of interest to divide the lens population into red and blue (or early- and late-type) sub-samples and to focus mainly on the early type galaxies, since they will not be so susceptible  to inclination effects \citep{novak}. Without color or morphology information this was a difficult task with our data, but by selecting a sample of galaxies with an axis ratio, b$/$a, between 0.5 and 0.8 we are selecting mostly early type galaxies \citep{alam}. The results from applying this cut in axis ratio to our lenses is shown in Figure \ref{fig:ourshape3}. Using preferentially early type galaxies yields a shear ratio of 0.61$\pm$0.14.  The favors a halo with ellipticity of $\sim$0.4 and represent a more significant detection of non-spherical halos than when using the entire lens sample. It is important to note that in this analysis we make the assumption that the galaxies we are using are isolated, so a comparison to MOND is perhaps not completely justified. However, the lensing signal in this analysis is measured on scales dominated by the contribution to the "1-halo" term from the matter distribution around the galaxy itself rather than from the host halo of any satellite galaxies within groups or clusters.

A recent analysis of isolated galaxies in the SDSS indicated that the number density of satellite galaxies is higher near the major axes of a central galaxies than the minor axes \citep{brainerd05}. This, together with the suggestion that satellites show a preference for radial alignment with their hosts \citep{agustsson}, indicates that the measured flattening from galaxy-galaxy lensing may be underestimating the true halo flattening. The signal along the major axes will be suppressed by the intrinsic alignment of the satellites, thus our measurements should be considered a lower limit on halo ellipticity.  In order to quantify this effect it would be interesting to measure the preferred alignment of satellite galaxies, and its dependence on their position angle with respect to their host galaxy for a large sample of hosts and satellites. Galaxy-galaxy lensing measurements using redshift information can eliminate this potential contaminant as well as identify a sample of isolated foreground galaxies to eliminate the influence of the group or cluster environment.

\section{Summary}

We have used early $i^\prime$ data from the CFHTLS to detect a significant galaxy-galaxy lensing signal, and to constrain the velocity dispersion of galaxy halos at a redshift of 0.3. The measured velocity dispersion for an L$_*$ galaxy is 137$\pm$11 km s$^{-1}$ (corresponding to circular velocity V$_*$=194$\pm15$km s$^{-1}$). This result is consistent with previous galaxy-galaxy lensing estimates \citep{klein, henk05}.

In addition, we were able to estimate the total mass of an L$_*$ galaxy at redshift 0.3 to be 2.2$\pm 0.4$ $\times 10^{12}$ $h^{-1} $M$_\odot$, which corresponds to a virial mass, M$_{200}$ of  1.1$\pm 0.2$ $\times 10^{12}$ $h^{-1} $M$_\odot$. This is in good agreement with the results from RCS \citep{henk05}, SDSS \citep{mandelbaum06c}, and GEMS \citep{heymans06}.

We also saw some evidence for non-spherical dark matter halos, though a definitive answer awaits more data. With multi-color data we will be able to repeat the anisotropic weak lensing measurements for sub-samples of lenses selected by color, morphology and redshift, This will also allow a detailed comparison to the low redshift results obtained by Mandelbaum et al. (2006a).

Galaxy-galaxy lensing is a unique and powerful tool for studying the dark matter halos of galaxies to large projected distances. This technique has now been successfully applied to many different data sets, but with planned and ongoing major surveys the results will only improve. Any complete theory of galaxy formation must be able to to properly map both the baryonic and dark matter components of galaxies, and understanding which galaxies live in which halos is a major piece of this puzzle. Many observations can help in understanding the baryonic component of galaxies, while simulations are best at explaining the dark matter. Gravitational lensing has an important role to play in uniting simulations and other observations by connecting galaxies to their halos.

There is clearly much more that can and will be done with the entire CFHTLS data set. In particular, having photometric redshifts will aid the analysis immensely. At present, we have lenses and sources over a wide range of redshifts and are estimating shear in angular bins which mix together many physical scales. Furthermore, photometric redshifts will permit a clear division between lens and source populations, rather than a statistical one based upon observed magnitudes as used here. This should help alleviate the concerns of intrinsic alignments. Redshift information also permits the division of lenses into different redshift bins, thus allowing the study of halo evolution.

A further use of the photometric redshifts could be to create catalogues of isolated lens galaxies. If the lens sample includes both central galaxies and their satellites, and the satellites are aligned with their host, then the galaxy-galaxy lensing signal could be biased. Also, the lensing signal around satellites will be influenced both by the dark matter halo of the satellite and the central galaxy. This issue can be addressed by measuring the lensing around only around the isolated ``central'' galaxies or by attempting to measure the signal around the central and satellite galaxies separately \citep[e.g.,][]{henk05,yang}. It is also possible to model the satellite contribution to the lensing signal  \citep[e.g.,][]{guzik}.

We also plan to measure the morphologies of all of the lens galaxies so that we can divide the sample into early and late-types and try to detect differences in their halo masses, shapes and M/L ratios. Furthermore, with the complete data set we should have sufficient statistical power to be able to distinguish between dark matter profiles. For example, we should be able to tell whether the tangential shear is best fit by an NFW profile \citep{nfw}, an isothermal sphere or a Moore profile \citep{moore}.

\acknowledgments

We would like to thank Joel Primack and Tereasa Brainerd for helpful discussions. The authors would also like to thank the anonymous referee for a careful reading of the manuscript and many helpful suggestions that improved the paper.

This work is based on observations obtained with MegaPrime/MegaCam, a joint project of CFHT and CEA/DAPNIA, at the Canada-France-Hawaii Telescope (CFHT) which is operated by the National Research Council (NRC) of Canada, the Institut National des Science de l'Univers of the Centre National de la Recherche Scientifique (CNRS) of France, and the University of Hawaii. This work is based in part on data products produced at TERAPIX and the Canadian Astronomy Data Centre as part of the Canada-France-Hawaii Telescope Legacy Survey, a collaborative project of NRC and CNRS.


\begin{thebibliography}{}


\bibitem[{{Agustsson} \& {Brainerd}(2006)}]{agustsson1}
{Agustsson}, I. \& {Brainerd}, T.~G. 2006, \apjl, 644, L25

\bibitem[{{Agustsson} \& {Brainerd}(2006)}]{agustsson}
{Agustsson}, I. \& {Brainerd}, T.~G. 2006, \apj, 650, 550

\bibitem[{{Alam} \& {Ryden}(2002)}]{alam}
{Alam}, S.~M.~K. \& {Ryden}, B.~S. 2002, \apj, 570, 610

\bibitem[{{Allgood} {et~al.}(2006){Allgood}, {Flores}, {Primack}, {Kravtsov},
  {Wechsler}, {Faltenbacher}, \& {Bullock}}]{allgood}
{Allgood}, B., {Flores}, R.~A., {Primack}, J.~R., {Kravtsov}, A.~V.,
  {Wechsler}, R.~H., {Faltenbacher}, A., \& {Bullock}, J.~S. 2006, \mnras, 367,
  1781

\bibitem[{{Astier} {et~al.}(2006){Astier}, {Guy}, {Regnault}, {Pain},
  {Aubourg}, {Balam}, {Basa}, {Carlberg}, {Fabbro}, {Fouchez}, {Hook},
  {Howell}, {Lafoux}, {Neill}, {Palanque-Delabrouille}, {Perrett}, {Pritchet},
  {Rich}, {Sullivan}, {Taillet}, {Aldering}, {Antilogus}, {Arsenijevic},
  {Balland}, {Baumont}, {Bronder}, {Courtois}, {Ellis}, {Filiol}, {Gon{\c
  c}alves}, {Goobar}, {Guide}, {Hardin}, {Lusset}, {Lidman}, {McMahon},
  {Mouchet}, {Mourao}, {Perlmutter}, {Ripoche}, {Tao}, \&
  {Walton}}]{cfhtlsdeep}
{Astier}, P., {et~al.}  2006, \aap, 447, 31

\bibitem[{{Bartelmann}(2006)}]{bart}
Bartelmann, M. 1996, \aa, 313, 697

\bibitem[{{Bekenstein}(2004)}]{bek}Bekenstein, J.~D. 2004, \prd, 70, 1

\bibitem[{{Bertin} \& {Arnouts}(1996)}]{bertin}
{Bertin}, E. \& {Arnouts}, S. 1996, \aaps, 117, 393

\bibitem[{Brainerd {et~al.}(1996)Brainerd, Blandford, \& Smail}]{bbs}
Brainerd, T., Blandford, R., \& Smail, I. 19c96, \apj, 466, 623

\bibitem[{{Brainerd}(2005)}]{brainerd05}
{Brainerd}, T.~G. 2005, \apjl, 628, L101

\bibitem[{{Brainerd \& Wright}(2000)}]{brainerd}
Brainerd, T.~G. \& Wright, C.~O. 2000, e-print astro-ph/0006281

\bibitem[{{Brown} {et~al.}(2003){Brown}, {Taylor}, {Bacon}, {Gray}, {Dye},
  {Meisenheimer}, \& {Wolf}}]{brown}
{Brown}, M.~L., {Taylor}, A.~N., {Bacon}, D.~J., {Gray}, M.~E., {Dye}, S.,
  {Meisenheimer}, K., \& {Wolf}, C. 2003, \mnras, 341, 100

\bibitem[\protect\astroncite{Bullock et~al.}{2001}]{bullock}
Bullock, J.~S., Kolatt, T.~S., Sigad, Y., Somerville, R.~S., Kravtsov, A.~V.,
  Klypin, A.~A., Primack, J.~R., and Dekel, A. 2001,
 \mnras,  321, 559

\bibitem[{{Cohen} {et~al.}(2000){Cohen}, {Hogg}, {Blandford}, {Cowie}, {Hu},
  {Songaila}, {Shopbell}, \& {Richberg}}]{cohen}
{Cohen}, J.~G., {Hogg}, D.~W., {Blandford}, R., {Cowie}, L.~L., {Hu}, E.,
  {Songaila}, A., {Shopbell}, P., \& {Richberg}, K. 2000, \apj, 538, 29

\bibitem[{{Conroy} {et~al.}(2005){Conroy}, {Coil}, {White}, {Newman}, {Yan},
  {Cooper}, {Gerke}, {Davis}, \& {Koo}}]{conroy}
{Conroy}, C., {Coil}, A.~L., {White}, M., {Newman}, J.~A., {Yan}, R., {Cooper},
  M.~C., {Gerke}, B.~F., {Davis}, M., \& {Koo}, D.~C. 2005, \apj, 635, 990

\bibitem[{{Dav{\'e}} {et~al.}(2001){Dav{\'e}}, {Spergel}, {Steinhardt}, \&
  {Wandelt}}]{dave}
{Dav{\'e}}, R., {Spergel}, D.~N., {Steinhardt}, P.~J., \& {Wandelt}, B.~D.
  2001, \apj, 547, 574

\bibitem[{{Dubinski} \& {Carlberg}(1991)}]{dubinski}
{Dubinski}, J. \& {Carlberg}, R.~G. 1991, \apj, 378, 496

\bibitem[{{Faber} \& {Jackson}(1976)}]{fb}
{Faber}, S.~M. \& {Jackson}, R.~E. 1976, \apj, 204, 668

\bibitem[{{Fern{\'a}ndez-Soto} {et~al.}(1999){Fern{\'a}ndez-Soto}, {Lanzetta},
  \& {Yahil}}]{fern}
{Fern{\'a}ndez-Soto}, A., {Lanzetta}, K.~M., \& {Yahil}, A. 1999, \apj, 513, 34

\bibitem[{{Frei} \& {Gunn}(1994)}]{frei}
{Frei}, Z. \& {Gunn}, J.~E. 1994, \aj, 108, 1476

\bibitem[{{Guzik} \& {Seljak}(2001)}]{guzik}{Guzik}, J. and {Seljak}, U. 2001, \mnras, 321, 439 

\bibitem[{{Heymans} {et~al.}(2006{\natexlab{a}}){Heymans}, {Van Waerbeke}, {Bacon}, {Berge},
  {Bernstein}, {Bertin}, {Bridle}, {Brown}, {Clowe}, {Dahle}, {Erben}, {Gray},
  {Hetterscheidt}, {Hoekstra}, {Hudelot}, {Jarvis}, {Kuijken}, {Margoniner},
  {Massey}, {Mellier}, {Nakajima}, {Refregier}, {Rhodes}, {Schrabback}, \&
  {Wittman}}]{heymans}
{Heymans}, C., {et~al.} 2006{\natexlab{a}}, \mnras, 350, 1323

\bibitem[{{Heymans} {et~al.}(2006{\natexlab{b}}){Heymans et al.}}]{heymans06} {Heymans}, C., {et~al.} 2006{\natexlab{b}}, \mnras, 371, L60

\bibitem[{{Hoekstra {et~al.}}(1998)Hoekstra, Franx, Kuijken, \& Squires}]{henk98}
Hoekstra, H., Franx, M., Kuijken, K., \& Squires, G. 1998, \apj, 504, 636


\bibitem[{{Hoekstra {et~al.}}(2002)Hoekstra et al.}]{henk02}{Hoekstra}, H. and {Yee}, H.~K.~C. and {Gladders}, M.~D. and {Barrientos}, L.~F. and {Hall}, P.~B. and {Infante}, L. 2002 , \apj, 572, 55

\bibitem[{Hoekstra {et~al.}(2003)Hoekstra, Franx, Kuijken, Carlberg, \&
  Gladders}]{henk03}
Hoekstra, H., Franx, M., Kuijken, K., Carlberg, R., \& Gladders, H.~Y. 2003,
  \mnras, 340, 609

\bibitem[{Hoekstra {et~al.}(2004)Hoekstra, Yee, \& Gladders}]{henk04}
Hoekstra, H., Yee, H., \& Gladders, M.~D. 2004, \apj, 606, 67

\bibitem[{Hoekstra {et~al.}(2005)Hoekstra et al.}]{henk05} {Hoekstra}, H. and {Hsieh}, B.~C. and {Yee}, H.~K.~C. and {Lin}, H., \&{Gladders}, M.~D. 2005, \apj, 635, 73

\bibitem[{{Hoekstra} {et~al.}(2006){Hoekstra et al.}}]{cfhtlsshear}
{Hoekstra}, H., {Mellier}, Y., {van Waerbeke}, L., {Semboloni}, E., {Fu}, L.,
  {Hudson}, M.~J., {Parker}, L.~C., {Tereno}, I., \& {Benabed}, K. 2006, \apj, 647, 116

\bibitem[{Hudson {et~al.}(1998)Hudson, Gwyn, Dahle, \& Kaiser}]{hudson}
Hudson, M.~J., Gwyn, S. D.~J., Dahle, H., \& Kaiser, N. 1998, \apj, 503, 531

\bibitem[{{Ilbert} {et~al.}(2006){Ilbert}, {Arnouts}, {McCracken},
  {Bolzonella}, {Bertin}, {Le Fevre}, {Mellier}, {Zamorani}, {Pello}, {Iovino},
  {Tresse}, {Bottini}, {Garilli}, {Le Brun}, {Maccagni}, {Picat}, {Scaramella},
  {Scodeggio}, {Vettolani}, {Zanichelli}, {Adami}, {Bardelli}, {Cappi},
  {Charlot}, {Ciliegi}, {Contini}, {Cucciati}, {Foucaud}, {Franzetti},
  {Gavignaud}, {Guzzo}, {Marano}, {Marinoni}, {Mazure}, {Meneux}, {Merighi},
  {Paltani}, {Pollo}, {Pozzetti}, {Radovich}, {Zucca}, {Bondi}, {Bongiorno},
  {Busarello}, {De La Torre}, {Gregorini}, {Lamareille}, {Mathez}, {Merluzzi},
  {Ripepi}, {Rizzo}, \& {Vergani}}]{ilbert}
{Ilbert}, O., {et~al.} 2006, \aap, 457, 841
  
  
\bibitem[{Kaiser {et~al.}(1995)Kaiser, Squires, \& Broadhurst}]{ksb}
Kaiser, N., Squires, G., \& Broadhurst, T. 1995, \apj, 449, 460

\bibitem[{{Kleinheinrich} {et~al.}(2005){Kleinheinrich}, {Rix}, {Erben},
  {Schneider}, {Wolf}, {Schirmer}, {Meisenheimer}, {Borch}, {Dye}, {Kovacs}, \&
  {Wisotzki}}]{klein}
{Kleinheinrich}, M., {Rix}, H.-W., {Erben}, T., {Schneider}, P., {Wolf}, C.,
  {Schirmer}, M., {Meisenheimer}, K., {Borch}, A., {Dye}, S., {Kovacs}, Z., \&
  {Wisotzki}, L. 2005, \aap, 439, 513

\bibitem[{Lin {et~al.}(1999)Lin, Yee, Carlberg, Morris, Sawicki, Patton, Wirth,
  \& Shepherd}]{lin}
Lin, H., Yee, H. K.~C., Carlberg, R.~G., Morris, S.~L., Sawicki, M., Patton,
  D.~R., Wirth, G., \& Shepherd, C.~W. 1999, \apj, 518, 533
  
  \bibitem[{{Mandelbaum} {et~al.}(2006{\natexlab{a}}){Mandelbaum}, {Seljak},
  {Kauffmann}, {Hirata}, \& {Brinkmann}}]{mandelbaum06c}
{Mandelbaum}, R., {Seljak}, U., {Kauffmann}, G., {Hirata}, C.~M., \& {Brinkmann},
  J. 2006{\natexlab{a}}, \mnras, 368, 715

\bibitem[{{Mandelbaum} {et~al.}(2006{\natexlab{b}}){Mandelbaum}, {Hirata},
  {Broderick}, {Seljak}, \& {Brinkmann}}]{mandelbaumshape}
{Mandelbaum}, R., {Hirata}, C.~M., {Broderick}, T., {Seljak}, U., \&
  {Brinkmann}, J. 2006{\natexlab{b}}, \mnras, 370, 1008

\bibitem[{{Massey} {et~al.}(2007)}]{step2}{Massey}, R., {et~al.} 2007, \mnras, 376, 13
	
\bibitem[{{Milgrom}(2002)}]{milgrom2}
{Milgrom}, M. 2002, \nat, 46, 741

\bibitem[{Moore {et~al.}(1999)Moore, Ghigna, Governato, Quinn, Stadel, \&
  Tozzi}]{moore}
Moore, B., Ghigna, S., Governato, F., Quinn, G. L.~T., Stadel, J., \& Tozzi, P.
  1999, \apj, 524, L19

\bibitem[{{Mortlock} \& {Turner}(2001)}]{mortlock}
{Mortlock}, D.~J. \& {Turner}, E.~L. 2001, \mnras, 327, 557

\bibitem[{Navarro {et~al.}(1996)Navarro, Frenk, \& White}]{nfw}
Navarro, J.~F., Frenk, C.~S., \& White, S.~D. 1996, \apj, 462, 563

\bibitem[{{Novak} {et~al.}(2006){Novak}, {Cox}, {Primack}, {Jonsson}, \&
  {Dekel}}]{novak}
{Novak}, G.~S., {Cox}, T.~J., {Primack}, J.~R., {Jonsson}, P., \& {Dekel}, A.
  2006, \apjl, 646, L9
  
\bibitem[{{Prada} {et~al.}(2003){Prada}, {Vitvitska}, {Klypin}, {Holtzman},
  {Schlegel}, {Grebel}, {Rix}, {Brinkmann}, {McKay}, \& {Csabai}}]{prada}
{Prada}, F., {Vitvitska}, M., {Klypin}, A., {Holtzman}, J.~A., {Schlegel},
  D.~J., {Grebel}, E.~K., {Rix}, H.-W., {Brinkmann}, J., {McKay}, T.~A., \&
  {Csabai}, I. 2003, \apj, 598, 260

\bibitem[{{Schneider} \& {Rix}(1997)}]{rix}
{Schneider}, P. \& {Rix}, H.-W. 1997, \apj, 474, 25

\bibitem[{Sheldon {et~al.}(2004)}]{sheldongg}
Sheldon, E.~S. {et~al.} 2004, \aj, 127, 2544

\bibitem[{{Sheth} {et~al.}(2003){Sheth}, {Bernardi}, {Schechter}, {Burles},
  {Eisenstein}, {Finkbeiner}, {Frieman}, {Lupton}, {Schlegel}, {Subbarao},
  {Shimasaku}, {Bahcall}, {Brinkmann}, \& {Ivezi{\'c}}}]{sheth}
{Sheth}, R.~K., {Bernardi}, M., {Schechter}, P.~L., {Burles}, S., {Eisenstein},
  D.~J., {Finkbeiner}, D.~P., {Frieman}, J., {Lupton}, R.~H., {Schlegel},
  D.~J., {Subbarao}, M., {Shimasaku}, K., {Bahcall}, N.~A., {Brinkmann}, J., \&
  {Ivezi{\'c}}, {\v Z}. 2003, \apj, 594, 225

\bibitem[{{Skordis} {et~al.}(2006){Skordis}, {Mota}, {Ferreira}, \&
  {B{\oe}hm}}]{skordis}
{Skordis}, C., {Mota}, D.~F., {Ferreira}, P.~G., \& {B{\oe}hm}, C. 2006, \prl, 96, 011301

\bibitem[{{Springel} {et~al.}(2004){Springel}, {White}, \&
  {Hernquist}}]{springel04}
{Springel}, V., {White}, S.~D.~M., \& {Hernquist}, L. 2004, in IAU Symposium, The shapes of simulated dark matter halos, ed S. D. Ryder, D. J. Pisano, M. A. Walker, and K. C. Freeman (San Francisco: ASP), 421

\bibitem[\protect\astroncite{{van Waerbeke} et~al.}{2006}]{vw06}
{van Waerbeke}, L., {White}, M., {Hoekstra}, H., and {Heymans}, C.: 2006, Astroparticle Physics, 26, 91

\bibitem[\protect\astroncite{Wright \& Brainerd}{2000}]{wrightbrain}
Wright, C.~O. \& Brainerd, T.~G.: 2000,
\newblock {\em \apj} {\bf 534}, 34


\bibitem[{{Yang} {et~al.}(2006)}]{yang}
{Yang}, X., {Mo}, H.~J., {van den Bosch}, F.~C., {Jing}, Y.~P., {Weinmann},
  S.~M., \& {Meneghetti}, M. 2006, \mnras, 373, 1159

\bibitem[{Yee {et~al.}}(2000)]{yee} Yee, H., Morris, S., Lin, H., Carlberg, R., Hall, P., Sawicki, M., Patton, D.,
  Wirth, G., Ellingson, E., \& Shepherd, C. 2000, \apjs, 187, 425

\bibitem[{Yoo {et~al.}}(2006)]{yoo} Yoo, J., Tinker, J.~L., Weinberg, D.~H., Zheng, Z., Katz, N.,  \& Dav\'e, R., 2006, \apj, 652, 26

\bibitem[{{Zaritsky} \& {White}(1994)}]{zaritsky}
{Zaritsky}, D. \& {White}, S.~D.~M. 1994, \apj, 435, 599

\bibitem[{{Zehavi} {et~al.}(2004){Zehavi}, {Weinberg}, {Zheng}, {Berlind},
  {Frieman}, {Scoccimarro}, {Sheth}, {Blanton}, {Tegmark}, {Mo}, {Bahcall},
  {Brinkmann}, {Burles}, {Csabai}, {Fukugita}, {Gunn}, {Lamb}, {Loveday},
  {Lupton}, {Meiksin}, {Munn}, {Nichol}, {Schlegel}, {Schneider}, {SubbaRao},
  {Szalay}, {Uomoto}, \& {York}}]{zehavi}
{Zehavi}, I., {et~al.} 2004,
  \apj, 608, 16

\bibitem[{{Zhao} {et~al.}(2006){Zhao}, {Bacon}, {Taylor}, \& {Horne}}]{zhao}
{Zhao}, H., {Bacon}, D.~J., {Taylor}, A.~N., \& {Horne}, K. 2006, \mnras, 368, 171

\end{thebibliography}
\end{document}